\documentclass
[notitlepage,showpacs,a4paper,nobalancelastpage,twocolumn,prb]{revtex4-1}%
\usepackage{amsfonts}
\usepackage{amsmath}
\usepackage{amssymb}
\usepackage{graphicx,epstopdf}
\usepackage{appendix}
\usepackage{array}
\usepackage{color}
\usepackage[normalem]{ulem}
\usepackage{CJK}
\providecommand{\U}[1]{\protect\rule{.1in}{.1in}}

\newcommand\rmv{\bgroup\markoverwith {\textcolor{red}{\rule[0.5ex]{2pt}{0.4pt}}}\ULon}

\newcommand{\fudan}{\affiliation{Department of Physics and State Key Laboratory of Surface Physics, Fudan University, Shanghai 200433, China}}

\newcommand{\nn}{\nonumber\\}
\newcommand{\ov}{\over}

\def\strutdepth{\dp\strutbox}
\newcommand\marginhack{\strut\vadjust{\kern-\strutdepth\hacksize}}
\newcommand\hacksize{\vtop to \strutdepth{ \baselineskip\strutdepth \vss\llap{{\tiny {\margintext}\quad }}\null}}
\newcommand\margintext{margin texts big and wide and wide }

\newcommand{\Eq}[1]{Eq.~\eqref{#1}}

\newcommand{\midb}[1]{\left[ #1 \right]}
\newcommand{\smlb}[1]{\left( #1 \right)}

\newcommand{\ket}[1]{\left|#1\right\rangle}

\newcommand{\half}{{1\over 2}}

\newcommand{\bk}{\ensuremath{\mathbf k}}

\newcommand{\bq}{\ensuremath{\mathbf q}}

\newcommand{\bB}{\ensuremath{\mathbf B}}

\newcommand{\bR}{\ensuremath{\mathbf R}}
\newcommand{\bS}{\ensuremath{\mathbf S}}

\begin{document}
\begin{CJK*}{UTF8}{gbsn} 
\title{Dynamical spin-spin coupling of quantum dots}
\author{Vahram L. Grigoryan}
\author{Jiang Xiao (萧江)}
\email[Corresponding author:~]{xiaojiang@fudan.edu.cn}
\fudan

\begin{abstract}
We carried out a nested Schrieffer-Wolff transformation of an Anderson
two-impurity Hamiltonian to study the spin-spin coupling between two dynamical
quantum dots under the influence of rotating transverse magnetic field. As a result of the rotating field, we predict a novel Ising type spin-spin coupling mechanism between quantum dots, whose strength is tunable via the magnitude of the rotating field. The strength of the new coupling could be comparable to the strength of the RKKY coupling. The dynamical coupling with the intristic RKKY coupling enables to construct a four level system of maximally entangled Bell states in a controllable manner.
\end{abstract}

\maketitle
\end{CJK*}

Quantum control of electron spins in semiconductor nanostructures is a central
issue in the emerging fields of spintronics and quantum information
processing. Electron spins confined in a semiconductor quantum dot (QD) was
proposed \cite{Loss} as a qubit for the realization of scalable quantum
computers. In the context of quantum-computational applications it is
necessary to couple qubits which are not nearest neighbors. The long-range
type spin-spin interactions include the Ruderman-Kittel-Kasuya-Yosida (RKKY)
interaction, \cite{Ruderman} Anderson's superexchange interaction
\cite{Anderson} and couplings mediated by cavity photons, \cite{Imamoglu} etc.

In all of the above mentioned systems the localized spins in the
QDs are assumed to be static. The dynamics of mesoscopic systems
have been discussed in context of charge pumping. \cite{Thouless}
This mechanism is called quantum pumping and was experimentally
realized first by Pothier \textit{et. al.} using QD in 1990. \cite{Pothier}
The pumping mechanism is also suitable for producing spin
currents, an essential ingredient for spintronics.
\cite{Wolf,Zutic} A pure spin pump was theoretically proposed by
Mucciolo et al \cite{Mucciolo} in 2001 and was experimentally
realized by the Marcus group. \cite{Watson}

The spin-pumping induced dynamic exchange coupling between ferromagnetic films
separated by normal-metal spacers is reported by experiments with sufficiently
large normal spacers. \cite{Heinrich} The dynamical coupling was first
discussed in the context of electron spin resonance by Barnes
in 1974, \cite{Barnes} who pointed out its long range nature as compared to
static coupling. In the context of ferromagnetic resonance experiments,
dynamic exchange coupling has been widely studied by different authors.
\cite{Hurdequint,Barnes,Heinrich,Lenz,Tserkovnyak} It was shown that in
multilayers and superlattices, on top of the equilibrium spin currents that
communicate the nonlocal static exchange coupling, a dynamic exchange
interaction with a much longer range becomes important.

Despite the extensive study of dynamical coupling in ferromagnetic
nanostructures, \cite{Hurdequint,Barnes,Heinrich,Lenz,Tserkovnyak}
the quantum counterpart of the phenomenon has not been discussed
yet. In this paper we study the dynamical coupling of two QDs in
Fermi sea. We find a new type of coupling Hamiltonian between the spins in two QDs, which is due to the dynamical cotunneling process induced by the rotating transverse magnetic field applied. The strength of this new dynamical coupling is tunable via the magnitude of the transverse magnetic field, and found to be comparable to the static RKKY coupling. The eigen energies with this dynamical coupling is different from that of the RKKY coupling, and can be used identifying the dynamical coupling strength. The corresponding eigen states, a set of Bell states, can be used for quantum computation.

 We consider two singly occupied QDs residing
in an electron bath and are
exposed to a magnetic field with weak DC component $B_{\parallel}$ along $Z$ axis (which
breaks the spin degeneracy by a Zeeman splitting) and a strong AC\ component $\bB_{\perp}\left(
t\right)$ in the $XY$ plane whose frequency
satisfies the resonance condition of the QD. \cite{Engel} The Hamiltonian of
the 2-QD system can be written as%

\begin{equation}
H=H_{0}+H_{t}+H_{B}\left(  t\right)  ,
\label{eq1}
\end{equation}
where%
\begin{subequations}
\begin{align}
H_{0}&=\sum_{i,\sigma}\varepsilon_{i}^{\sigma}n_{i}^{\sigma}+\sum_{i}U_{i}%
n_{i}^{\uparrow}n_{i}^{\downarrow}+\sum_{k,\sigma}\epsilon_{k}c_{k}^{\sigma\dag}c_{k}^{\sigma},
\label{eq2a}\\
H_{t}&=\sum_{i,k,\sigma}\left[  T_{i,k}\left(  \bR_{i}\right)  d_{i}^{\sigma\dag
}c_{k}^{\sigma}+T_{i,k}^{\ast}\left(  \bR_{i}\right)  c_{k}^{\sigma\dag} d_{i}^{\sigma}\right]  ,
\label{eq2b}\\
H_{B}\left(  t\right)  &=\sum_{i}\hbar \omega_{\bot}\smlb{  d_{i}^{\uparrow\dag}
d_{i}^{\downarrow}e^{-i\omega t}+d_{i}^{\downarrow\dag}d_{i}^{\uparrow}e^{i\omega t}}  .
\label{eq2c}
\end{align}
\end{subequations}
$H_{0}$ is the energy of non-interacting quantum dots and
conduction electrons where $d_{i}^{\sigma\dag}\left(
d_{i}^{\sigma}\right)  $ creates (annihilates) an electron with
spin $\sigma=\pm1\left(  \uparrow ,\downarrow\right)$ in $ $\ QD-$i$
$\left(  i=1,2\right)  $. $\varepsilon
_{i}^{\sigma}=\varepsilon_{i}+\sigma\hbar \omega_{\parallel}/{2}$ is the energy for spin-$\sigma$ in QD-$i$ with $\hbar \omega_{\parallel}=g\mu _{B}B_{\parallel}$ the Zeeman splitting due to external magnetic field $B_{\parallel}$ and gyromagnetic factor $g$ and Bohr magneton $\mu_{B}$.
$n_{i}^{\sigma}\equiv d_{i}^{\sigma\dag}d_{i}^{\sigma}$ is the
number operator of QD-$i$. $U_{i}$ is the Coulomb interaction energy on QD-$i$. The third term in Eq. $\left(  \ref{eq2a}\right)  $ stands
for the kinetic energy for the noninteracting electrons in the bath with $c_{k}^{\sigma}$
being the annihilation operator of a conduction electron with
momentum $k$ and spin $\sigma$ with energy $\epsilon_{k}.$ $H_{t}$\ is the tunneling
Hamiltonian between the localized electrons in QD-$i$ and the
conduction electrons with the tunneling rates at the QD position
$\bR_{i},$ $T_{i,k}\left(
\bR_{i}\right)  =T_{i,k}e^{-i\bk\cdot\bR_{i}}.%
$\cite{Tzen,Cesar} The effect of a rotating transverse magnetic field $\bB_{\perp}\left(
t\right)  =B_{\perp}\left[  \boldsymbol{i}\cos\left(  \omega t\right)
+\boldsymbol{j}\sin\left(\omega t\right)  \right]$ is in $H_{B}\left(  t\right)  $\ with $\hbar\omega_{\bot}=g\mu_{B}B_{\perp}$ and driving frequency $\omega.$ We assume that the QDs are in the Kondo regime, \textit{i.e.} $\varepsilon_{i}<\epsilon_{k}<U_{i}$ and the transfer matrix elements
$T_{i,k}\left(  \bR_{i}\right)  $ between the dots and the
continuum are small compared with $\epsilon_{k}$ and $U_{i},$ \textit{i.e.} $T_{i,k}\ll
U_{i},U_{i}-\epsilon_{k},\epsilon_{k}$. Under this conditions the number of electrons on the dot is a well-defined quantity. To eliminate the explicit time dependence of the Hamiltonian (\ref{eq1}) we make a unitary tranfromtaion to the frame of reference, rotating with frequency $\omega.$

 Following Refs.
\onlinecite{Kolley1,Minh,Kolley2,Braun}, we use a two-stage or nested
Schrieffer-Wolff (SW) \cite{Schrieffer} transformation to derive an
effective spin Hamiltonian to obtain the low-energy spin
interactions of the system. The perturbative tunneling Hamiltonian $H_{t} $ \cite{Caroli} enables us to apply
the SW transformation to the total Hamiltonian $H$ to
eliminate $H_{t}$:%
\begin{equation}
H'=e^{S}He^{-S}=H+\left[  S,H\right]  +\frac{1}{2}\left[  S,\left[
S,H\right]  \right]  +\cdot\cdot\cdot,\label{eq3}%
\end{equation}
where the generator operator $S$ is required to satisfy $H_{t}+\left[
S,H_{0}\right]  =0:$%
\begin{equation}
S=\sum_{i,k,\sigma}\smlb{  \frac{n_{i}^{\bar{\sigma}}}{E_{ki}-U_{i}}+\frac{1-n_{i}^{\bar{\sigma}}}{E_{ki}}}  T_{i,k}^{\ast}c_{k}^{\sigma\dag}%
d_{i}^{\sigma}-H.c.\label{eq4}%
\end{equation}
with $E_{ki} \equiv \epsilon_k - \varepsilon_i^\sigma$.  
In the absence of magnetic field, the above transformation reduces to two impurity Anderson model. \cite{Ng}%
 We are interested in the subspace of single
occupancy of the QDs, \textit{i.e.} one requires
$n_{i}=\sum_{\sigma}n_{i}^{\sigma}=1.$ This constraint can be
established by using the Gutzwiller operator
$P=\prod\limits_{i}\left(
n_{i}^{\uparrow}-n_{i}^{\downarrow}\right)  ^{2},$ \cite{Kolley3}
 and we retain only the 
contributions in $H'$ that survive under the projection with
$PH'P\neq0$. Up to second order
of tunneling rate after
Gutzwiller projection%
\begin{equation}
H'=H_{0}+H_{B}+\frac{1}{2}\left[  S,H_{t}\right]  +\frac{1}{2}\left[
S,\left[  S,H_{B}\right]  \right]  +\cdot\cdot\cdot,\label{eq5}%
\end{equation}
where the third term is second order in tunneling rate and includes the Kondo (also called \textquotedblleft
s-d\textquotedblright) Hamiltonian plus a potential scattering term.
\cite{Schrieffer}

To study the low-energy spin interactions of the two-impurity Anderson model, a
nested \cite{Kolley1,Minh,Kolley2,Braun} or generalized \cite{Zhou,Tzen,Cesar}
SW transformation has been used to derive an effective spin interaction Hamiltonian. The
purpose of the second transformation is to remove (at least partially) the
contribution in second-order in tunneling rate, \textit{i.e.} the third term in Eq. $\left(  \ref{eq5}\right).  $ \cite{Kolley1} The second SW transformation
should be done with the generator operator $S_{2}^{t}$ fulfilling $\left[  S_{2}%
^{t},H_{0}\right]  =-\frac{1}{2}\left[  S,H_{t}\right]  $. Note that
$S_{2}^{t}\varpropto T_{i,k}^{2}.$ This transformation reveals 4-th order interactions such as RKKY and a correlated Kondo term. \cite{Kolley1,Tzen}

Using the same idea as above, we perform a different second SW transformation using generator operator $S_{2}^{B}$
\begin{equation}
H''=e^{S_{2}^{B}}H'e^{-S_{2}^{B}}\label{eq6}%
\end{equation}
to eliminate the second order interaction term $\frac{1}{2}\left[  S,\left[
S,H_{B}\right]  \right]  $ that corresponds to the rotating magnetic field in Eq. $\left(  \ref{eq5}\right)  ,$ which requires
$\left[  S_{2}^{B},H_{0}\right]  =-\frac{1}{2}\left[  S,\left[  S,H_{B}%
\right]  \right]  ,$ $\left(  S_{2}^{B}\varpropto T_{i,k}^{2}\right)  .$
The purpose of this transformation is to extract the interactions
induced by the external rotating magnetic field. The
generator operator $S_{2}^B$ fulfilling the above condition in the rotating
frame of reference with frequency $\omega$ is 
\begin{align}
S_{2}^{B}&=\frac{\hbar\omega_{\bot}}{2}\sum_{i,k,q}\frac{P_{i,k,q}}{\epsilon_{k}%
-\epsilon_{q}}\nonumber\\
&\times\left[  n_{k,q}\left(  S_{i}^{+}+S_{i}^{-}\right)  -n_{i}\left(
S_{k,q}^{+}+S_{k,q}^{-}\right)  \right]  .\label{eq7}%
\end{align}
with $S_{k,q}^{+\left(  -\right)  }=c_{k}^{\uparrow\left(  \downarrow\right)
\dag}c_{q}^{\downarrow\left(  \uparrow\right)  }$, $S_{i}^{+\left(  -\right)
}=d_{i}^{\uparrow\left(  \downarrow\right)  \dag}d_{i}^{\downarrow\left(
\uparrow\right)  },$ $n_{k,q}=\sum_{\sigma}c_{k}^{\sigma\dag}c_{q}^{\sigma}$
and
\begin{equation}
P_{i,k,q}= T_{i,k}^{\ast}T_{i,q}
\midb{\frac{1}{(E_{ki}-U_{i})(E_{qi}-U_{i})}-\frac{1}{E_{ki}E_{qi}}}  .\label{eq8}%
\end{equation}
Note that for small Zeeman field the expressions are simplified by assuming
that the band energy levels are not spin-dependent $\varepsilon_{i}^{\sigma
}=\varepsilon_{i}.$ After the transformation, the effective Hamiltonian
becomes%
\begin{align}
H''=H_{0}&+H_{B}+\frac{1}{2}\left[  S,H_{t}\right]  \nonumber\\
&+\frac{1}{2}\left[  S_{2}^{B},\left[  S,H_{t}\right]  \right]  +\frac{1}%
{4}\left[  S_{2}^{B},\left[  S,\left[  S,H_{B}\right]  \right]  \right].
\label{eq9}%
\end{align}
The aim of SW transformations is to eliminate perturbatively "old" terms from Hamiltonians in favor of "novel" interactions. The first SW transformation in Eq. $\left(\ref{eq5} \right )$ gives rise to the second order (in tunneling rates) processes, including the standard Kondo interaction, and an analog of the spin pumping and spin torque interaction between the QDs and the continuum. The second SW transformation in Eq. $\left( \ref{eq9} \right )$ gives rise to the fourth order processes resulting from the rotating magnetic field. We discuss the second order and forth order processes below.

 In addition to the standard Kondo
Hamiltonian, the first SW transformation yields the spin pumping
induced by the rotating field and spin torque due to the
absorption of the pumped spins (the last term in Eq. $\left(
\ref{eq5}\right)  $). In the rotating frame
\begin{align}
\half\left[  S,\left[  S,H_{B}\right]  \right]  &=
{\hbar\omega_{\bot}\ov 2}\sum_{i,k,q}P_{i,k,q}e^{i\left(  \bk-\bq\right)
\cdot\bR_{i}} \label{eq11} \\
&\times\left[  n_{i}\left(  S_{k,q}^{+}+S_{k,q}^{-}\right)  -n_{k,q}\left(
S_{i}^{+}+S_{i}^{-}\right)  \right]  \nonumber.
\end{align}

\begin{figure}[b]
\begin{center}
\includegraphics[
trim=0.000000in 0.000000in -0.001981in 0.000000in,
height=0.7in,
width=3.2396in
]%
{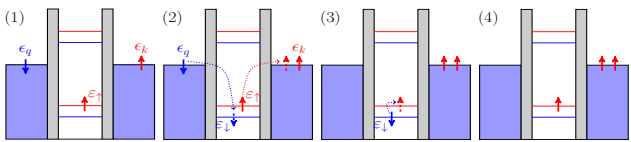}%
\caption{(Color online) Schematic plot of the cotunneling induced spin
pumping. The arrows correspond to the electrons with spin up and down. }%
\label{fig1}%
\end{center}
\end{figure}
The first term in Eq. $\left(  \ref{eq11}\right)  $ is responsible
for cotunneling induced spin pumping from QD to the continuum.
The second term gives the torque acting the QD by the pumped spins
in the continuum. The physics of the precession induced spin
pumping from the QD is depicted in Figure \ref{fig1}. Due to the
cotunneling processes, one spin-$\downarrow$ electron with
momentum $q$ and energy $\epsilon_{q}$ in the continuum tunnels
into the dot and occupies energy level $\varepsilon_{\downarrow}$,
followed by another spin-$\uparrow$ electron on
$\varepsilon_{\uparrow}$ tunneling out to the continuum with momentum $k$
and energy $\epsilon_{k}.$ Afterwards, due to the
rotating field the electron in the QD absorbs a "photon" and
transists to spin-$\uparrow.$ With these Kondo-type cotunneling processes, the spin in QD remains unchanged, but one spin down is flipped to spin up in the continuum via QD. Such processes keep
transferring electrons in the continuum from spin-$\downarrow$ to
spin-$\uparrow$. The reverse process
also happens if a spin-$\downarrow$ electron resides in the QD (not shown in the figure): spin-$\uparrow$ electron tunnels into the QD followed by the spin-$\downarrow$ electron tunneling out to the continuum, emits
a "photon" and flips to spin-$\downarrow.$ The average outcome is that the electrons
from continuum flows toward the scattering region, change their
spins and flow away from it. It should be noted, that as the
average number of pumped electrons with opposite spins are equal
and thus no spin or charge current is induced. But we are, in the
end, interested in fourth order processes, when spins pumped from
one QD can be absorbed by the second one and thus mediate new
coupling mechanism
between two dynamic QDs.%

 After two-step SW transformation,
the Hamiltonian in Eq. $\left(  \ref{eq9}\right)$ contains two fourth order terms, \textit{i.e.} the last two terms. The explicit form of fourth term is:
\begin{equation}
{1\ov2}\left[  S_{2}^{B},\left[  S,H_{t}\right]  \right]  =
{\hbar\omega_{\bot}J_{1}\ov2}\left[  n_{1}\left(  S_{2}^{+}+S_{2}^{-}\right)
+n_{2}\left(  S_{1}^{+}+S_{1}^{-}\right)  \right]  ,\label{eq12}%
\end{equation}
with
\begin{align*}
J_{1}=\frac{1}{4}\sum_{k,q}\sum_{\left\langle i,j\right\rangle }%
\frac{n_{k}-n_{q}}{\epsilon_{k}-\epsilon_{q}}P_{i,k,q}J_{j,q,k}e^{i\left(
\bk-\bq\right)  \cdot\bR_{i,j}},
\end{align*}
\begin{equation}
J_{i,k,q}= T_{i,k}^{\ast}T_{i,q}
\left(  \frac{1}{E_{ki}}+\frac{1}{E_{qi}}-\frac{1}{E_{ki}-U_{i}}-\frac
{1}{E_{qi}-U_{i}}\right)  .\label{eq13}
\end{equation}
and $\bR_{i,j}=\bR_{i}-\bR_{j}$.
The Eq. $\left(  \ref{eq12}\right)  $ represents a new effective coupling of
QDs induced by the spin flip scattering of electrons. The physics of new coupling can be understood as follows: the conduction electron tunnels into one
QD, flips its spin, and then tunnels out back into the continuum. Afterwards the flipped spin flows toward the
other QD and exchanges with the electron spin in it. The
exchange of electron spins between two QDs is via the RKKY-type coupling where the
energy of the intermediate excitation is given by $\epsilon_{q}-\epsilon_{k}%
$.\ It can be seen from Eq. $\left(  \ref{eq12}\right)  $ that spin-flip
scattering induced coupling reduces to an effective magnetic field from
adjacent quantum dot.

The last term of Eq. $\left(  \ref{eq9}\right)  $ %
\begin{align}
&\frac{1}{4}\left[  S_{2}^{B},\left[  S,\left[  S,H_{B}\right]  \right]
\right]  =\frac{\left (\hbar \omega_{\bot}\right)^{2}J_{2}}{4}\nonumber\\
&\times\left(  n_{1}n_{2}+S_{1}^{+}S_{2}^{+}+S_{1}^{-}S_{2}^{+}+S_{1}^{+}%
S_{2}^{-}+S_{1}^{-}S_{2}^{-}\right)  , \label{eq1.4}%
\end{align}
with 
\begin{align*}
J_{2}\equiv\frac{1}{2}\sum_{k,q}\sum_{\left\langle i,j\right\rangle
}\frac{n_{k}-n_{q}}{\epsilon_{k}-\epsilon_{q}}P_{i,k,q}P_{j,q,k}e^{i\left(
\bk-\bq\right)  \cdot\bR_{i,j}}.
\end{align*}
The Hamiltonian Eq. $\left(  \ref{eq1.4}\right)  $ describes the processes when
electrons after spin-flipping in one QD tunnel into the second one and flip again.

Since we are interested in the subspace of single occupancy of the
QDs $\left(  n_{i}=1\right)  $ we can extract the interactions that survive with Gutzwiller projection. The effective Hamiltonian in the laboratory frame becomes%
\begin{align}
H_{B}''(t)  &=\sum_{i}\hbar \omega_{\parallel} S_{i,z} +\sum_{i}g\mu_{B}\left(  2+J_{1}\right)
\bB_{\perp}\left(  t\right)  \cdot\bS_{i} \nn
&+\left(  g\mu_{B}\right)  ^{2}J_{2}
\midb{\bB_{\perp}\left(  t\right) \cdot\bS_{1}}
\midb{\bB_{\perp}\left(  t\right) \cdot\bS_{2}}.
\label{eq15}
\end{align}
This effective coupling between QDs due to the rotating magnetic field is the main result of this paper. The $J_1$ part in the second term in \Eq{eq15} represents that the spin $\bS_1$ of th QD-1 feels the magnetic field acting on $\bS_2$ on QD-2, and vice versa. While the third term of \Eq{eq15} represents the Ising-type coupling between the spins on QD-1 and QD-2 induced by the rotating magnetic field $\bB_{\perp}(t)$, and it magnitude is tunable by changing the magnitude of $\bB_{\perp}(t)$.
 We can perform a unitary transformation $\hat{U} \left(t\right)=e^{-i \sum_{i}{\omega S_{i,z}}}$ that removes the time dependence:
\begin{align}
H_{B}''  &=\hat{U} \left(t\right)^\dag \left( H_{B}''(t) - i \hbar \partial_{t} \right) \hat{U} \left(t\right) = \sum_{i}{\hbar \left( \omega_\parallel-\omega \right) S_{i,z}}\nn
&+\sum_{i}\hbar\omega_{\bot}\left(  2+J_{1}\right)
S_{i,x}+\left(\hbar\omega_{\bot}\right)^{2}J_{2}S_{1,x}S_{2,x}.
\label{eq.16}
\end{align}%
When the resonance condition $\left( \omega_\parallel=\omega \right)$ is fulfilled, \Eq{eq.16} becomes
\begin{equation}
H_{B}''=\sum_{i}\hbar\omega_{\bot}\left(  2+J_{1}\right)
S_{i,x}+\left(\hbar\omega_{\bot}\right)^{2}J_{2}S_{1,x}S_{2,x},\label{eq17}%
\end{equation}
where the first is the sum of Eq. $\left ( \ref{eq12}\right)$ and the external transverse magnetic field $H_{B}=\sum_{i}2\hbar\omega_{\bot}S_{i,x}$ in rotating reference frame and the second term corresponds to Eq. $\left ( \ref{eq1.4} \right)$ with $S_{i,x}=\frac{1}{2}\left(
S_{i}^{+}+S_{i}^{-}\right)  .$

 In Ref. \onlinecite{Coqblin} Coqblin and
Schrieffer presented their widely used approach
\cite{Schlottmann,Bazhanov,Yang,Coleman} to the two-impurity Anderson model.
After a single SW transformation and treating that Hamiltonian in second order
perturbation theory, they compute a RKKY-like spin-spin
interaction \cite{Ruderman} of the form $J_0\bS_{1}%
\cdot\bS_{2}$ with coupling constant $J_0=\sum_{k,q}J_{1,k,q}J_{2,q,k}%
\frac{n_{k}-n_{q}}{\epsilon_{k}-\epsilon_{q}}.$ We evaluate $\hbar\omega_{\bot
}J_{1}$ and $\left(\hbar\omega_{\bot}\right)^{2}J_{2}$ in Eq. $\left(  \ref{eq15}\right)  $
comparing them with RKKY coupling constant. For simplicity we assume identical
QDs $\left(  \varepsilon_{1}=\varepsilon_{2}\equiv\varepsilon,\text{ }%
U_{1}=U_{2}\equiv U\right)  .$ In symmetric Kondo regime the typical energy of
QDs $\left(  \varepsilon=0\right)  $ is $\left\vert \epsilon_{k,q}\right\vert
\equiv\epsilon=U/2\simeq0.1$ meV. \cite{Goldhaber} By setting the rotating
field strength $B_{\perp}=1$ T and $g=2,$ we have from Eqs. $\left(
\ref{eq8}\right)  $ and $\left(  \ref{eq13}\right)  $ $\left\vert \hbar\omega
_{\bot}J_{1}\right\vert /J_0\simeq\hbar\omega_{\bot}/4\epsilon=0.28$ and
$\left(\hbar\omega_{\bot}\right)^{2}J_{2}/J_0\simeq\left(\hbar\omega_{\bot}\right)^{2}/4\epsilon^{2}=0.32$. 
This means the magnitude of the new coupling terms in \Eq{eq17} is comparable to the RKKY coupling.

\begin{table}[t]
\centering
\begin{tabular}{c|c|l} 
State & Wave function & Energy \\ \hline\hline
$\ket{1}$ & $\Phi^{+} + \Psi^{+}$ & $J_0+2 \hbar\omega_\perp \smlb{2+J_1}+\smlb{\hbar\omega_\perp}^2 J_2$  \\
$\ket{2}$ & $\Phi^{+} - \Psi^{+}$ & $J_0-2 \hbar\omega_\perp \smlb{2+J_1}+\smlb{\hbar\omega_\perp}^2 J_2$ \\
$\ket{3}$ & $\Phi^{-}$ & $J_0-\smlb{\hbar\omega_\perp}^2 J_2$ \\
$\ket{4}$ & $\Psi^{-}$ 		& $-3J_0-\smlb{\hbar\omega_\perp}^2 J_2$
\end{tabular}
\caption{Eigen-energies and eigen-states for $H_{\rm eff}$ in \Eq{eqn:Heff}. } 
\label{tab:Epsi}
\end{table}

\begin{figure}[t]
\includegraphics[width=.8\columnwidth]{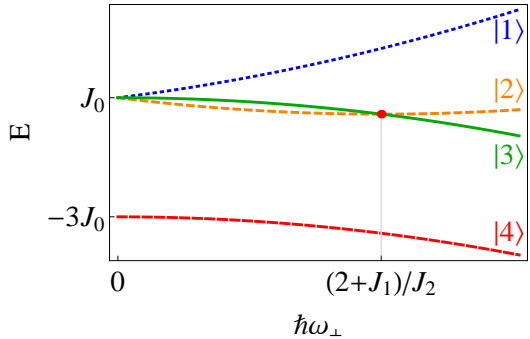}%
\caption{(Color online) The eigenenergies of the eigenstates $\ket{1}, \ket{2},\ket{3}$ and $\ket{4}$ as a function of the magnitude of the transverse magnetic field in units of $J_0.$}%
\label{fig2}%
\end{figure}

We study the effective Hamiltonian (Eq. \ref{eq15}) in the rotating reference frame in the presence of RKKY coupling $J_0\bS_{1}\cdot\bS_{2}$, which always exists in the system. The total effective Hamiltonian becomes 
\begin{equation}
\label{eqn:Heff}
H_{\rm eff} = H''_B + J_0 \bS_{1} \cdot \bS_{2}.
\end{equation}
Using the four Bell states: $\Phi^{\pm}=\frac{1}{\sqrt{2}} \left( \ket{\downarrow \downarrow} \pm \ket{ \uparrow \uparrow} \right)$ and $\Psi^{\pm}=\frac{1}{\sqrt{2}} \left( \ket{\downarrow \uparrow} \pm \ket{ \uparrow \downarrow} \right)$ as basis, the eigen-energies and the eigen-states are listed in Table \ref{tab:Epsi}.
In Figure \ref{fig2}, we plot the eigenenergies of the effective Hamiltonian $H_{\rm eff}$ as a function of the magnitude of the transverse magnetic field $B_\perp = \hbar\omega_\perp/g\mu_B$. It is seen that in the limit of no transverse field, we have singlet and (threefold degenerate) triplet states. The transverse magnetic field removes the degeneracy of the triplet states and enables to construct a four level system in a controllable manner, in which two of the states ($\ket{2}$ and $\ket{3}$) swap positions at $\hbar\omega_\perp = (2+J_1)/J_2$. This feature can be used for quantum computation in desirable set of maximally entangled Bell states. Finally, the transverse magnetic field dependence of the eigen energies in Figure \ref{fig2} can be measured in experiment, and the coupling strengths $J_{1}$ and $J_{2}$ can be inferred from the slope (at zero field) and curvature of the lines, respectively.

In conclusion, 
 using a 2-stage SW
transformation, we transform a two-impurity Anderson model into an effective spin Hamiltonian. Without the rotating magnetic field the second order expansion yields the standard Kondo Hamiltonian
for two impurities with additional scattering terms. \cite{Schrieffer,Kolley1,Tzen,Cesar,Kolley3} The
introduction of the rotating magnetic field gives rise to a magnetic field-induced spin pumping from QD and the torque that experiences the QD from the continuum via the cotunneling processes. 
These cotunneling processes yield to two additional QD coupling mechanisms: 1) the QD feels a non-local magnetic field acting on the neighboring QD; 2) the QDs are coupled via an Ising-like coupling. Because of its dynamical origin from the rotating field, the new coupling mechanism is intrinsically different from all existing static coupling mechanisms such as RKKY coupling. More importantly, the strength of the new dynamical coupling is tunable via the magnitude of the transverse magnetic field, and found to be comparable to the RKKY coupling strength at reasonably large rotating magnetic field. The interplay of the RKKY and the new dynamical coupling enables to construct a four level system of the maximally entangled Bell states in a controllable manner.

 This work was supported by the special funds for the Major State Basic
Research Project of China (No. 2011CB925601) and the National Natural Science
Foundation of China (Grants No. 11004036 and No. 91121002).

\end{document}